\documentclass[letter]{aa}
\usepackage{graphicx}
\usepackage{amsmath}
\usepackage{geometry}
\usepackage[varg]{txfonts}

\begin{document}

\title{Electrons under the dominant action of shock-electric fields} 

\titlerunning{Electrons under the dominant action of shock-electric fields}
\authorrunning{Fahr \& Verscharen}

\author{Hans J. Fahr\inst{1}
\and Daniel Verscharen\inst{2}} 

\offprints{Hans J. Fahr, \email{hfahr@astro.uni-bonn.de}}

\institute{Argelander Institute for Astronomy, University of Bonn, Auf dem H\"{u}gel 71, 53121 Bonn, Germany
  \and Space Science Center, University of New Hampshire, 8 College Road, Durham, NH 03824, USA}

\abstract{We consider a fast magnetosonic multifluid shock as a
representation of the solar-wind termination shock. We assume the
action of the transition happens in a three-step process: In the first
step, the upstream supersonic solar-wind plasma is subject to a strong
electric field that flashes up on a small distance scale $\Delta z\simeq U_1/
\Omega _{\mathrm e}$ (first part of the transition layer), where $\Omega_{\mathrm e}$ is the electron gyro-frequency and $U_1$ is the upstream speed. This electric field both
decelerates the supersonic ion flow and accelerates the electrons up to
high velocities. In this part of the transition region, the
electric forces connected with the deceleration of the ion flow strongly dominate over the Lorentz forces. We, therefore, call this part the demagnetization region. In the second phase, Lorentz forces due to convected magnetic fields 
compete with the electric field, and the highly anisotropic and energetic electron distribution
function is converted into a shell distribution with energetic shell
electrons storing about 3/4 of the upstream ion kinetic energy. In the third phase, the plasma particles thermalize due to the relaxation of free energy by plasma instabilities. The first part of the transition region opens up a new thermodynamic
degree of freedom never before taken into account for the electrons, since
the electrons are usually considered to be enslaved to follow the
behavior of the protons in all velocity moments like density, bulk velocity,
and temperature. We show that electrons may be the
downstream plasma fluid that dominates the downstream plasma pressure.     }


\maketitle

\section{Introduction}

Recent literature on astrophysical shocks concludes with increasing conviction that a shock, such as the solar-wind termination shock, is a phenomenon in which multifluid effects play an important role. This means that, for the understanding of  global shock physics,  effects arising from different plasma species must be taken into account while all species have to fulfill the Rankine-Hugoniot conservation laws concertedly. For instance, previous studies indicate that pickup ions substantially modify the structure of the solar-wind termination shock. 
In addition,  pickup ions are generally considered an important fluid component since they transport a major fraction of the entropized upstream kinetic energy in the form of thermal energy into the downstream regime of the shock \citep[see][]{decker08}. \citet{zank10} and
\citet{fahr07,fahr10,fahr11} found relations between the upstream and downstream ion distribution functions that are different for solar-wind protons and for pickup protons, respectively.
However, these refinements did not lead to a fully satisfying representation of some properties of the shocked plasma observed by Voyager 2 \citep{richardson08}. 
In fact, \citet{chalov13} demonstrate that it is necessary to allow for solar-wind electrons to behave like a third, independent plasma fluid  to achieve an agreement between the model results and the plasma data presented by \citet{richardson08}. 
Their parameter study shows that the best-fit results for the observed proton temperature are then achieved if the electrons are assumed to be heated to a temperature that is higher than the downstream proton temperature by a factor $\gtrsim 10$.
We refer to the works of \citet{leroy84}, \citet{tokar86}, and \citet{schwartz88} on electron heating at fast-mode shocks. 
Also, plasma shock simulations, in which electrons are treated kinetically, show preferential shock heating of the plasma electrons \citep[see][]{lembege03,lembege04}. Two-stream and viscous interactions lead to a demagnetization of the electrons; i.e., they are not Lorentz-wound by the magnetic fields and attain an increase in their temperature by a factor of 50 or more on the downstream side of the shock.
In a different approach, \citet{leroy82} and \citet{goodrich84} predict that electrons carry out perpendicular electric drifts that are different from those of the ions due to the shock-electric field, thereby establishing an electric current that modifies the surface-parallel magnetic field.

With a similar level of consistency,  \citet{fahr12} describe
the conditions of the upstream and downstream plasma in bulk-frame
systems with frozen-in magnetic fields. They treat the transition from the upstream side to the downstream side of the shock as an instantaneous kinetic reaction in the velocity distribution function via the Liouville-Vlasov theorem, predicting all of the relevant downstream plasma quantities. In this model, a semikinetic representation of the multifluid termination shock with mass- and charge-specific reactions of protons and electrons to the electric shock ramp leads to excessive electron heating as described by \citet{fahr12} or \citet{fahr13}.
Based on these studies, we assume that electrons enter the downstream side as a strongly heated, massless plasma fluid that dominates the downstream plasma pressure.
A recently published multifluid termination shock reconstruction by \citet{zieger15}, which describes the electrons as a separate, independently reacting fluid, supports this notion.
This study finds that electrons are preferentially heated with respect to ions, converting overshoot velocities into thermal energy supporting the claims made by \citet{chalov13}, \citet{fahr15}, and \citet{fahr15b}. Furthermore, the study finds that the partial pressure of the heated electrons plays a dominant role in the downstream plasma flow.

In the following study, we investigate more carefully for what reason and in which way the plasma electrons  behave as a preferentially heated fluid at the shock crossing.

\section{The shock electric field and charge separation}

Strongly space-dependent electric and magnetic fields are typical for the transition region of astrophysical shocks, and none of the conventional test-particle descriptions, such as the expansion of the particle motion into different forms of electromagnetic drifts, apply under these conditions.
This is certainly the case when electric forces locally flash up and become strongly dominant over Lorentz forces so that the latter forces can be safely neglected for the first-order motion analysis. In this section, we look into the details of this special situation that arises at shocks such as the solar wind-termination shock.

\begin{figure}
  \resizebox{\hsize}{!}{\includegraphics{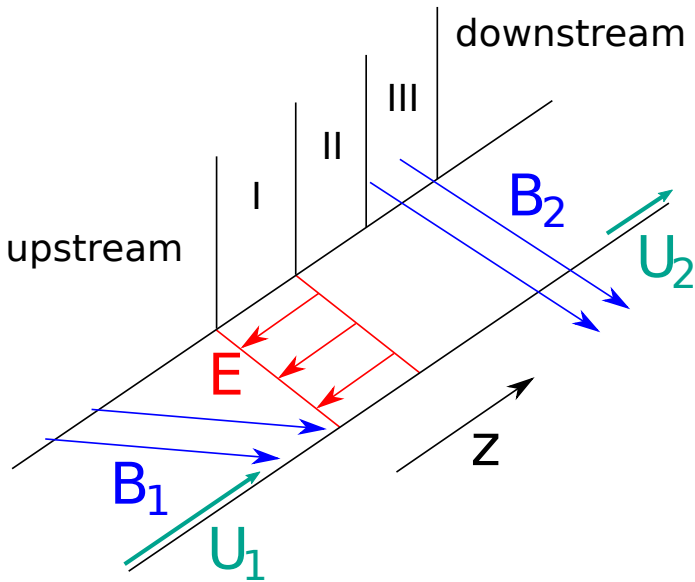}}
  \caption{Illustration of the three-phase shock. Phase I: electric forces dominate the deceleration of the plasma (demagnetization region); Phase II: magnetic Lorentz forces dominate; Phase III: quasi-equilibrium relaxation of the plasma and thermalization through plasma instabilities.}
  \label{sketch}
\end{figure}

We assume that the shock is characterized by a three-phase transition in space. In the first phase (the demagnetization region),  electric forces dominate the deceleration of the plasma. In the second phase, Lorentz forces dominate the deceleration over the electric forces. While in the first two phases the transition is isothermal for protons, the third phase on the downstream side represents the region in which the particle thermalization due to streaming plasma instabilities takes place. We focus on the first phase of the shock transition (the demagnetization region). We illustrate the three-phase shock in Fig.~\ref{sketch}.

We first estimate the typical field conditions at a shock of this nature. In a sufficiently small surface element of the termination shock, the curvature of the shock surface is negligible. We approximate this surface element under consideration as a planar surface and, consequently, assume a one-dimensional geometry of the system. In this approximation, the shock parameters only depend on the coordinate $z$ parallel to the shock normal.
We assume a representation of the ion bulk-flow velocity $U$ as a function of  $z$ of the form 
\begin{equation}\label{bulk}
U(z)=U_{1}-\frac{1}{2}\Delta U \left[ 1+\tanh \left(\frac{ z}{\alpha }\right)\right]
\end{equation}
where $\Delta U=U_{1}-U_{2}$. Indexes  1 and 2 characterize the asymptotic
upstream and downstream conditions of the flow, respectively. The quantity $\alpha $
determines the transition length at the shock. We choose the coordinate system in which $\vec B$ lies in the x-z-plane.
In addition, we assume that the $\vec B$-field is frozen into the main momentum flow  (i.e,. the ion flow). Therefore, the frozen-in condition under steady-state conditions, $\partial \vec{B}/\partial t=\nabla\times [\vec{U}\times \vec{B}]=0$,  yields
\begin{equation}
\vec{B}=\vec{B}_{x}(z)+\vec{B}_{z}
\end{equation}
in our one-dimensional coordinate system, where $B_{z}=B_{z1}=\mathrm{const.}$ and $B_{x}=B_{x1} U_{1}/U(z)$. 

The $z$-component of the ion momentum equation is written as
\begin{equation}\label{momentum}
m_{\mathrm p}U\frac{\mathrm dU}{\mathrm dz}=-\frac{1}{n_{\mathrm p}}\frac{\mathrm dP_{\mathrm p}}{\mathrm dz} +eE,
\end{equation}
where $m_{\mathrm p}$ is the proton mass $e$ is the elementary charge, $E$ is the electric field that is parallel to the flow direction $\hat{\vec e}_z$, $P_{\mathrm p}$ is the proton pressure, and $n_{\mathrm p}$ is the proton density  \citep[see also][]{verscharen08,fahr13}.
In our isothermal approximation ($T_{\mathrm p}=\mathrm{const.}$) for the first phase of the shock transition, we find for the pressure-gradient term in the momentum equation by applying the flux-conservation requirement  $\Phi _{\mathrm e}=\Phi _{\mathrm p}=n_{\mathrm p}U=\mathrm{const}$ that
\begin{equation}
-\frac{1}{n_{\mathrm p}}\frac{\mathrm dP_{\mathrm p}}{\mathrm dz}=\frac{k_{\mathrm B}T_{\mathrm p}}{U}\frac{\mathrm dU}{\mathrm dz},
\end{equation}
where $k_{\mathrm B}$ is the Boltzmann constant.
This leads to the following expression for the electric field 
\begin{equation}\label{Ecompl}
E=\frac{m_{\mathrm p}}{e}U\frac{\mathrm dU}{\mathrm dz}\left(1-\frac{1}{2}\frac{c_{\mathrm p}^2}{U^2}\right),
\end{equation}
where $c_{\mathrm p}=\sqrt{2k_{\mathrm B}T_{\mathrm p}/m_{\mathrm p}}$ is the ion thermal speed.
The correction term $c_{\mathrm p}^2/U^2$ on the right-hand side is small compared to 1 for high-Mach-number shocks in the first phase of the shock transition, which justifies neglecting the pressure gradient.
With Eq.~(\ref{bulk}), the electric field as a function of $z$ is given by
\begin{equation}
E=-\frac{m_{\mathrm p}}{2e\alpha}U\,\Delta U\mathrm{sech}^2\left(\frac{z}{\alpha}\right).
\end{equation}

The electric field $E$ is self-consistently generated by the charge separation between protons and electrons according to Gauss' law:
\begin{equation}
\frac{\mathrm dE}{\mathrm dz}=4\pi e\left(n_{\mathrm p}-n_{\mathrm e}\right).
\end{equation}

\section{Dominance of electric forces over Lorentz forces}\label{electric}

We can assume that, as long as the electric force is much stronger than
the Lorentz force on the electrons, the motion of a single electron with  electron velocity $v_{\mathrm ez}$ corresponds to a permanent linear acceleration according to
\begin{equation}
m_{\mathrm e}v_{\mathrm ez}\frac{\mathrm d v_{\mathrm ez}}{\mathrm dz}=-eE.
\end{equation}
Integrating leads to
\begin{equation}
\left. \frac{1}{2}m_{\mathrm e}v_{\mathrm ez}^{2}\right| _{-\Delta }^{+\Delta
}=-\int\limits _{-\Delta }^{+\Delta }eE\,\mathrm dz.
\end{equation}
We assume that the initial energy of an electron with a velocity equal to the bulk-flow velocity on the upstream side, $\epsilon _{\mathrm e0}=m_{\mathrm e}U_{1}^{2}/2$, is negligible compared to the resulting downstream energy. We note that protons and electrons have equal upstream bulk velocities $U_1$ as well as equal upstream densities $n_{1}$. With this assumption and Eq.~(\ref{Ecompl}) in the limit $c_{\mathrm p}^2/U^2\ll 1$, we find for the $z$-dependent kinetic energy of the electron that
\begin{multline}\label{DSE}
\frac{1}{2}m_{\mathrm e}v_{\mathrm ez}^{2}=-\int\limits _{-\infty }^{z}eE(z^{\prime})\mathrm dz^{\prime}=\frac{m_{\mathrm p}}{2\alpha}\Delta U\int\limits_{-\infty}^{z}U(z^{\prime})\mathrm{sech}^2\left(\frac{z^{\prime}}{\alpha}\right)\mathrm dz^{\prime}\\
=\frac{1}{2}m_{\mathrm p}\Delta U\left\{\left(U_1-\frac{\Delta U}{2}\right)\left[1+\tanh\left(\frac{z}{\alpha}\right)\right]+\frac{\Delta U}{4}\mathrm{sech}^2\left(\frac{z}{\alpha}\right)\right\}.
\end{multline}
For the downstream electron kinetic energy, we obtain with Eq.~(\ref{DSE}) that
\begin{multline}
\frac{1}{2}m_{\mathrm e}v_{\mathrm ez}^{2}=\frac{m_{\mathrm p}}{2\alpha}\Delta U\int\limits_{-\infty}^{+\infty}U(z)\mathrm{sech}^2\left(\frac{z}{\alpha}\right)\mathrm dz\\
= \frac{1}{2}m_{\mathrm p} U_1^2\left(\frac{s^2-1}{s^2}\right)=0.84\frac{1}{2}m_{\mathrm p}U_{1}^2,
\end{multline}
with the observed compression ratio $s=U_{1}/U_{2}\approx 2.5$
\citep[see][]{richardson08}.
Therefore, the shock electrons may consume about 84\% of the upstream proton kinetic energy, which corresponds to about 0.7~keV or $v_{\mathrm ez2}\approx  10^{9}\,\mathrm{cm/s}$, respectively.

We choose an electron with $v_{\mathrm ez1}=U_1$ and follow its trajectory through the shock. This electron fulfills $v_{\mathrm ez}(z)=U_{\mathrm e}(z)$ throughout its transit. 
With this condition and flux conservation, $n_{\mathrm e}U_{\mathrm e}=\mathrm{const.}$, we directly determine the upstream-to-downstream density ratio for the electrons as
\begin{equation}
\frac{n_{\mathrm e2}}{n_1}=\frac{U_1}{U_{\mathrm e2}}=\sqrt{\frac{m_{\mathrm e}}{m_{\mathrm p}}}\sqrt{\frac{s^2}{s^2-1}}=0.025,
\end{equation}
which translates to $U_{\mathrm e2}=40U_{1}$.

\begin{figure}
  \resizebox{\hsize}{!}{\includegraphics{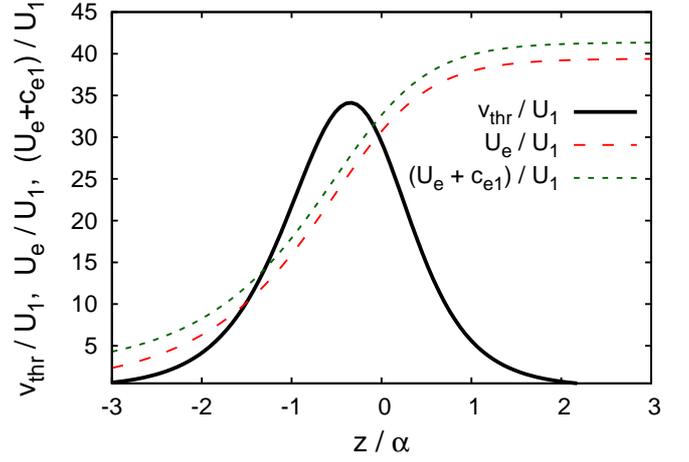}}
  \caption{Electron bulk velocity $U_{\mathrm e}$ from Eq.~(\ref{DSE}), threshold velocity $v_{\mathrm{thr}}$ from Eq.~(\ref{thr}), and $U_{\mathrm e}+c_{\mathrm e 1}$, where $c_{\mathrm e1}=\sqrt{2k_{\mathrm B}T_{\mathrm e1}/m_{\mathrm e}}$ is the upstream electron thermal speed, as functions of the spatial coordinate $z$. We use the following parameters: $s=2.5$, $U_1=4\times 10^7\,\mathrm{cm/s}$, $B_{x1}=2\times 10^{-7}\,\mathrm G$, $\alpha=10^8\,\mathrm{cm}$, and $T_{\mathrm e1}=2\times 10^4\,\mathrm K$.}
  \label{lorentz_cond}
\end{figure}

The electric acceleration of
the electron, however, only dominates compared to the acceleration due to Lorentz forces as long as 
\begin{equation}\label{Edom}
\left|eE\right|\geq \left|\frac{e}{c}v_{\mathrm ez}B_{x}\right|.
\end{equation}
This condition limits the resulting velocity to
\begin{equation}
\left|v_{\mathrm ez}\right|\leq \left| \frac{c}{B_{x}}E\right|=\left| \frac{1}{2\alpha} \frac{\Delta U}{U_1}\frac{m_{\mathrm p}c}{eB_{x1}}U^2\mathrm{sech}^2\left(\frac{z}{\alpha}\right)\right|, 
\end{equation}
which we write in normalized units as
\begin{equation}\label{thr}
\left|\frac{v_{\mathrm ez}}{U_1}\right|\le\left|\frac{1}{2}\frac{U_1}{\Omega_{\mathrm px1}\alpha}\left(\frac{U}{U_1}\right)^2\frac{\Delta U}{U_1}\mathrm{sech}^2\left(\frac{z}{\alpha}\right)\right|,
\end{equation}
where $\Omega_{\mathrm px1}=eB_{x1}/m_{\mathrm p}c$. We define the normalized threshold velocity $v_{\mathrm{thr}}/U_1$ as the right-hand side of Eq.~(\ref{thr}).
Beyond this limit, Lorentz forces begin to compete with electric forces and need to be taken into account. We show this condition in Fig.~\ref{lorentz_cond}.
The black solid line shows the threshold velocity $v_{\mathrm {thr}}$ in units of the upstream bulk velocity $U_1$.  Considering again an electron with $v_{\mathrm ez}(z)\approx U_{\mathrm e}(z)$, we determine the profile of $U_{\mathrm e}$ as a function of $z$ from Eq.~(\ref{DSE}). We show this profile as the red dashed line in Fig~\ref{lorentz_cond}.  The bulk velocity fulfills the condition in Eq.~(\ref{thr}) in the range between $z\approx-1.5\alpha$ and $z\approx -0.1\alpha$. We also plot the sum of the electron bulk velocity and the upstream electron thermal speed as the green dashed line. This line indicates the region in which the majority of the solar-wind electrons fulfill Eq.~(\ref{thr}). This condition is fulfilled between $z\approx -1.2\alpha$ and $z\approx -0.1\alpha$. As these considerations show, the demagnetization region is restricted to the region $z<0$.

In the limit of a quasi-parallel shock, $\Omega_{\mathrm px1}\rightarrow 0$ which leads to $v_{\mathrm{thr}}\rightarrow\infty$. Therefore, the quasi-parallel shock is solely a result of the electrostatic interaction while Lorentz forces are negligible. On the other hand, a quasi-perpendicular shock still has a finite $v_{\mathrm{thr}}$, which can be $>U_{\mathrm e}$ depending on the finite shock parameters $U_1$, $\Omega_{\mathrm px1}$, and $\alpha$.

\section{Motion under first-order Lorentz forces}\label{motion}

We assume that, within a finite region, the electric forces are strongly dominant according to Eq.~(\ref{Edom}).
The zeroth-order motion of an electron is then determined by
\begin{equation}
\frac{\mathrm d\vec v_{\mathrm e}^{0}}{\mathrm dt}=-\frac{e}{m_{\mathrm e}} \vec E,
\end{equation}
while the first-order Lorentz force is given by
\begin{equation}
\vec{K}_{\mathrm L}^{1}=-\frac{e}{c}\left[\vec{v}_{\mathrm e}^{0}\times \vec{B}\right].
\end{equation}

Under the assumption that the electric field is purely oriented along the $z$-direction, we obtain 
\begin{equation}
{v_{\mathrm e}^{0}}^2(z)-{v_{\mathrm e1}^{0}}^{2}=-2\int\limits_{-\infty}^{z}\frac{e}{m_{\mathrm e}}E(z^{\prime})\,\mathrm dz^{\prime}.
\end{equation}
We furthermore assume that the upstream electron velocity $v_{\mathrm e1}^{0}$
is negligible.
Now we allow for a first-order velocity component in the $y$-direction (i.e., the $\hat{\vec e}_z\times \hat{\vec{b}}$-direction, where $\hat{\vec b}=\vec B/|\vec B|$) due to the first-order Lorentz force $K_{\mathrm L}^1$: 
\begin{equation}\label{KL}
v_{\mathrm ey}^{1}=\int\limits_{-\infty}^z \frac{1}{m_{\mathrm e}}K_{\mathrm L}^{1}(z^{\prime})\frac{\mathrm dt}{\mathrm dz^{\prime}}\,\mathrm dz^{\prime},
\end{equation}
where $K_{\mathrm L}^1$ is given by
\begin{equation}
K_{\mathrm L}^{1}(s)=-\frac{e}{c}\left|\left[\vec{v}_{\mathrm e}^{0}(z)\times \vec{B}(z)\right]\right|=-\frac{e}{c}B_x(z)\sqrt{-\frac{2e}{m_{\mathrm e}}\int\limits_{-\infty}^{z}E(z^{\prime})\,\mathrm dz^{\prime}}.
\end{equation}
With Eq.~(\ref{KL}), we find
\begin{equation}
v_{\mathrm ey}^{1}=-\int\limits_{-\infty}^z \frac{e}{m_{\mathrm e}c}B_x(z^{\prime})\sqrt{-\frac{2e}{m_{\mathrm e}}\int\limits_{-\infty}^{z^{\prime}}E(z^{\prime\prime})\,\mathrm dz^{\prime\prime}}%
\frac{\mathrm dt}{\mathrm dz^{\prime}}\,\mathrm dz^{\prime}.
\end{equation}
Using $\mathrm dt/\mathrm dz=1/v_{\mathrm e}^{0}(z)$, we obtain the solution
\begin{equation}
v_{\mathrm ey}^{1}=-\int\limits_{-\infty}^{z}\frac{e}{m_{\mathrm e}c}B_x(z^{\prime})\,\mathrm dz^{\prime}=-\int\limits_{-\infty}^{z}\Omega_{\mathrm e x}(z^{\prime})\,\mathrm dz^{\prime}=-z\left\langle \Omega_{\mathrm ex}\right\rangle _{z},
\end{equation}
where $\Omega_{\mathrm ex}(z)$ denotes the local electron gyro-frequency based on $B_x$, and $\left\langle
\Omega_{\mathrm ex}\right\rangle _{z}$ is the average gyro-frequency in the integration domain.

It is noteworthy that the electron motion in the $y$-direction does not depend on the electric-field configuration in the transition region as long as the Lorentz force can be considered to be of higher
order than the electric force.
Therefore, the curling of the particle trajectory around the frozen-in field only happens after a time $\tau_{\mathrm{Le}}$ when the action of the Lorentz force changed the velocity by about its magnitude,
$(e/c)v_{\mathrm ez}B_{x} \tau _{\mathrm{Le}}\approx m_{\mathrm e}v_{\mathrm ez}$, yielding $\tau _{\mathrm{Le}}\approx m_{\mathrm e}c/eB_x=1/\Omega _{\mathrm ex}$. During this period, the electron moves in the $z$-direction by the amount $\Delta z=\tau_{\mathrm{Le}}v_{\mathrm ez}=v_{\mathrm ez}/\Omega _{\mathrm ex}$.
With our results for $v_{\mathrm ez}$ from Sect.~\ref{electric}, we find that the permitted extent of the overall transition region can roughly be
estimated as $\Delta z\approx v_{\mathrm ez}/\Omega _{\mathrm ex}\approx  10^{8}\,\mathrm{cm}$.

\section{Discussion and conclusions}

In this paper, we study the action of shock-electric fields on
electrons and ions entering from the upstream regime of the shock into the shock
transition region. We split the latter into three consecutive regions:
the first is demagnetized and operates like an electric double layer
where the action of the upflashing shock-electric field strongly dominates
over Lorentz forces. In the second region, Lorentz forces due to the piled-up
magnetic fields compete with or even dominate compared to electric forces. In the third phase, the overshooting energy is transferred into proton and electron heating by the action of kinetic plasma instabilities.

In Sect.~\ref{electric}, we determine the electron velocity profile. The downstream electron velocity before entering phase three of the shock is about 40 times the upstream velocity, while the downstream electron density is about 0.025 the upstream density. The shocked plasma opens up a new thermodynamic degree
of freedom allowing that a substantial fraction (about 84 percent) of the
upstream ion kinetic energy gets stored in the overshoot velocities of the
electrons which, in the second region of the shock transition, is deposited into a shell in velocity space
centered around the downstream ion bulk velocity. In that case, a major fraction of the upstream ion bulk-flow energy is available to be transferred into electron thermal energy.
We determine the threshold velocity $v_{\mathrm{thr}}$, which defines the separation between phases one and two of the shock transition. The electron motion is dominated by electric forces for all electrons with a velocity $<v_{\mathrm{thr}}$. For typical termination-shock parameters, we find that the region in which this condition is fulfilled for electrons with the electron bulk velocity extends from $z=-1.5\alpha$ to $z=-0.1\alpha$. 

In Sect.~\ref{motion}, we treat Lorentz-forces as a first-order correction to the plasma deceleration. We show that the first-order shock-perpendicular component of the electron velocity is determined by the average electron gyro-frequency in the shock layer. With this result, we determine the size of the demagnetization region (phase one) as $\Delta z\approx 10^8\,\mathrm{cm}$ for typical solar-wind parameters.

In earlier work  \citep[see][]{chalov13,fahr13,fahr15}, we emphasized the fact that the creation of energetic electrons at the plasma passage over the shock is
essential to fulfill the thermodynamic entropy requirements \citep{fahr15} and to arrive at downstream plasma properties that nicely fit the Voyager-2 measurements of \citet{richardson08}. \citet{zieger15} recently published a multifluid study of the solar-wind termination shock, which strongly supports our claim for the occurrence of energetic downstream
electrons by showing that the Voyager-2 measurements allow for a reasonably good theoretical fit with the model results only if the appearance of
energetic downstream electrons is taken into account.

The remaining question as to why these predicted energetic
electrons were not detected by the Voyager plasma analyzers,
can most probably be answered along the argumentation developed in a recent
paper by \citet{fahr15b}. These authors show that the energetic electrons create
a strongly increased electric charge-up of the spacecraft detectors and
thereby, owing to electric screening,  impede the detectors in measuring
countable fluxes of these electrons.       

We intend to continue the studies presented in this letter by kinetically analyzing the physics relevant for phases II and III of the three-phase shock.

\begin{acknowledgements}

We appreciate helpful discussions with Marty Lee. This work was supported in part by NSF/SHINE grant AGS-1460190.

\end{acknowledgements}

\bibliographystyle{aa}
\bibliography{shock_field_rev}

\begin{thebibliography}{20}
\expandafter\ifx\csname natexlab\endcsname\relax\def\natexlab#1{#1}\fi

\bibitem[{{Chalov} \& {Fahr}(2013)}]{chalov13}
{Chalov}, S.~V. \& {Fahr}, H.~J. 2013, \mnras, 433, L40

\bibitem[{{Decker} {et~al.}(2008){Decker}, {Krimigis}, {Roelof}, {Hill},
  {Armstrong}, {Gloeckler}, {Hamilton}, \& {Lanzerotti}}]{decker08}
{Decker}, R.~B., {Krimigis}, S.~M., {Roelof}, E.~C., {et~al.} 2008, \nat, 454,
  67

\bibitem[{{Fahr} {et~al.}(2015){Fahr}, {Richardson}, \& {Verscharen}}]{fahr15b}
{Fahr}, H.~J., {Richardson}, J.~D., \& {Verscharen}, D. 2015, \aap, 579, A18

\bibitem[{{Fahr} \& {Siewert}(2007)}]{fahr07}
{Fahr}, H.-J. \& {Siewert}, M. 2007, Astrophys.~Space Sci.~Transactions, 3, 21

\bibitem[{{Fahr} \& {Siewert}(2010)}]{fahr10}
{Fahr}, H.-J. \& {Siewert}, M. 2010, \aap, 512, A64

\bibitem[{{Fahr} \& {Siewert}(2011)}]{fahr11}
{Fahr}, H.-J. \& {Siewert}, M. 2011, \aap, 527, A125

\bibitem[{{Fahr} \& {Siewert}(2013)}]{fahr13}
{Fahr}, H.-J. \& {Siewert}, M. 2013, \aap, 558, A41

\bibitem[{{Fahr} \& {Siewert}(2015)}]{fahr15}
{Fahr}, H.-J. \& {Siewert}, M. 2015, \aap, 576, A100

\bibitem[{{Fahr} {et~al.}(2012){Fahr}, {Siewert}, \& {Chashei}}]{fahr12}
{Fahr}, H.-J., {Siewert}, M., \& {Chashei}, I. 2012, \apss, 341, 265

\bibitem[{{Goodrich} \& {Scudder}(1984)}]{goodrich84}
{Goodrich}, C.~C. \& {Scudder}, J.~D. 1984, \jgr, 89, 6654

\bibitem[{{Lemb{\`e}ge} {et~al.}(2004){Lemb{\`e}ge}, {Giacalone}, {Scholer},
  {Hada}, {Hoshino}, {Krasnoselskikh}, {Kucharek}, {Savoini}, \&
  {Terasawa}}]{lembege04}
{Lemb{\`e}ge}, B., {Giacalone}, J., {Scholer}, M., {et~al.} 2004, \ssr, 110,
  161

\bibitem[{{Lemb{\`e}ge} {et~al.}(2003){Lemb{\`e}ge}, {Savoini}, {Balikhin},
  {Walker}, \& {Krasnoselskikh}}]{lembege03}
{Lemb{\`e}ge}, B., {Savoini}, P., {Balikhin}, M., {Walker}, S., \&
  {Krasnoselskikh}, V. 2003, \jgr, 108, 1256

\bibitem[{{Leroy} \& {Mangeney}(1984)}]{leroy84}
{Leroy}, M.~M. \& {Mangeney}, A. 1984, Ann.~Geophys., 2, 449

\bibitem[{{Leroy} {et~al.}(1982){Leroy}, {Winske}, {Goodrich}, {Wu}, \&
  {Papadopoulos}}]{leroy82}
{Leroy}, M.~M., {Winske}, D., {Goodrich}, C.~C., {Wu}, C.~S., \&
  {Papadopoulos}, K. 1982, \jgr, 87, 5081

\bibitem[{{Richardson} {et~al.}(2008){Richardson}, {Kasper}, {Wang}, {Belcher},
  \& {Lazarus}}]{richardson08}
{Richardson}, J.~D., {Kasper}, J.~C., {Wang}, C., {Belcher}, J.~W., \&
  {Lazarus}, A.~J. 2008, \nat, 454, 63

\bibitem[{{Schwartz} {et~al.}(1988){Schwartz}, {Thomsen}, {Bame}, \&
  {Stansberry}}]{schwartz88}
{Schwartz}, S.~J., {Thomsen}, M.~F., {Bame}, S.~J., \& {Stansberry}, J. 1988,
  \jgr, 93, 12923

\bibitem[{{Tokar} {et~al.}(1986){Tokar}, {Aldrich}, {Forslund}, \&
  {Quest}}]{tokar86}
{Tokar}, R.~L., {Aldrich}, C.~H., {Forslund}, D.~W., \& {Quest}, K.~B. 1986,
  Phys.~Rev.~Lett., 56, 1059

\bibitem[{{Verscharen} \& {Fahr}(2008)}]{verscharen08}
{Verscharen}, D. \& {Fahr}, H.-J. 2008, \aap, 487, 723

\bibitem[{{Zank} {et~al.}(2010){Zank}, {Heerikhuisen}, {Pogorelov}, {Burrows},
  \& {McComas}}]{zank10}
{Zank}, G.~P., {Heerikhuisen}, J., {Pogorelov}, N.~V., {Burrows}, R., \&
  {McComas}, D. 2010, \apj, 708, 1092

\bibitem[{Zieger {et~al.}(2015)Zieger, Opher, Tóth, Decker, \&
  Richardson}]{zieger15}
Zieger, B., Opher, M., Tóth, G., Decker, R.~B., \& Richardson, J.~D. 2015,
  \jgr, n/a, 2015JA021437

\end{thebibliography}

\end{document}